\documentclass[letterpaper,11pt]{article}
\pdfoutput=1 

\usepackage{jheppub} 

\usepackage{dsfont}
\usepackage{color}

\def\ben{\begin{equation}}
\def\een{\end{equation}}
\def\hz{{\hat{z}}}

\title{Space-Time in the SYK Model}

\author{Sumit R. Das$^1$,}
\author{Animik Ghosh$^1$,}
\author{Antal Jevicki$^2$,}
\author{Kenta Suzuki$^2$}

\affiliation{$^1$Department of Physics and Astronomy, University of Kentucky, Lexington, KY 40506, U.S.A.}
\affiliation{$^2$Department of Physics, Brown University, 182 Hope Street, Providence, RI 02912, U.S.A.}

\emailAdd{das@pa.uky.edu}\emailAdd{animik.ghosh@uky.edu}
\emailAdd{antal\_jevicki@brown.edu}\emailAdd{kenta\_suzuki@brown.edu}

\abstract{
We consider the question of identifying the  bulk space-time of the SYK model. 
Focusing on the signature of emergent space-time of the (Euclidean) model,
we explain the need for non-local (Radon-type) transformations on external legs of $n$-point Green's functions.
This results in a dual theory with Euclidean AdS signature with additional leg-factors.
We speculate that these factors incorporate the coupling of additional bulk states similar to the discrete states of 2d string theory.}

\dedicated{{\large This work is dedicated to the memory of Joe Polchinski.}}

\begin{document}
\begin{flushright}
{UK/17-11} \\ {BROWN-HET-1728}
\end{flushright}

\maketitle
\flushbottom

\section{Introduction}
\label{sec:intro}
Detailed investigations of the Sachdev-Ye-Kitaev (SYK) model \cite{Sachdev:1992fk, Kitaev:2015, Kitaev:2014, Sachdev:2015efa, Polchinski:2016xgd, Maldacena:2016hyu, Jevicki:2016bwu, Jevicki:2016ito, Davison:2016ngz, Gross:2017hcz, Gross:2017aos, Kitaev:2017awl}
have given an interesting, highly nontrivial example of the AdS/CFT duality and a potential  framework for quantum black holes.
The model, which can be studied at Large $N$, features emergent reparametrization invariance at the IR critical point, with related out-of-time-order correlators exhibiting quantum chaos,
with a Lyapunov exponent characteristic of black holes \cite{Kitaev:2015}, providing the butterfly effect \cite{Shenker:2013pqa, Leichenauer:2014nxa, Shenker:2014cwa, Maldacena:2015waa, Polchinski:2015cea, Caputa:2016tgt, Anninos:2016szt, Turiaci:2016cvo}.
Random matrix theory interpretations have been studied in \cite{You:2016ldz, Garcia-Garcia:2016mno, Cotler:2016fpe, Hartnoll:2016mdv, Liu:2016rdi, Krishnan:2016bvg, Garcia-Garcia:2017pzl, Li:2017hdt, Kanazawa:2017dpd, Sonner:2017hxc}.
Related models have been studied also \cite{Danshita:2016xbo, Krishnan:2017lra, Erdmenger:2015xpq} with various generalizations \cite{Gross:2016kjj, Gu:2016oyy, Berkooz:2016cvq, Fu:2016vas, Fu:2016yrv, Garcia-Alvarez:2016wem, Nishinaka:2016nxg, Turiaci:2017zwd, Jian:2017unn, Chew:2017xuo, Murugan:2017eto, Yoon:2017gut, Peng:2017spg, Cai:2017vyk}.
The solution and properties  are shared with tensor type models \cite{Witten:2016iux, Gurau:2016lzk, Klebanov:2016xxf, Peng:2016mxj, Ferrari:2017ryl, Itoyama:2017emp, Narayan:2017qtw, Choudhury:2017tax, BenGeloun:2017jbi, Diaz:2017kub, deMelloKoch:2017bvv, Prakash:2017hwq, BenGeloun:2017vwn}.

For all these theories, bi-local observables, as proposed in \cite{Das:2003vw} in the context of $O(N)$ vector model/higher spin duality \cite{Klebanov:2002ja}, provide a route to a bulk construction of the dual theory with emergent space-time \cite{Koch:2010cy, Koch:2014mxa, Koch:2014aqa}. For the SYK model the IR and the near-IR limit are solvable, with evaluations \cite{Kitaev:2015, Maldacena:2016hyu, Jevicki:2016ito} of the invariant Schwarzian action representing the boundary Gravity degrees of freedom possibly related to JT type \cite{Teitelboim:1983ux, jackiw} dual theory
\cite{Almheiri:2014cka, Jensen:2016pah, Maldacena:2016upp, Engelsoy:2016xyb, Forste:2017kwy} 
(See also \cite{Cvetic:2016eiv, Hashimoto:2016dfz, Blake:2016jnn, Mandal:2017thl, Mezei:2017kmw, Taylor:2017dly, Grumiller:2017qao, Stanford:2017thb, Mertens:2017mtv,Sarosi:2017ykf}).

From symmetry considerations, the center of mass and relative coordinates of the two points in the bi-local fields can be interpreted as the coordinates of the Poincare patch of AdS$_2$ or dS$_2$ \cite{Polchinski:2016xgd, Maldacena:2016hyu, Jevicki:2016bwu} as in the simplest identification proposed in \cite{Das:2003vw}. Indeed, the action of fluctuations of the bi-local is a non-polynomial function of the AdS$_2$ Laplacian, indicating that the theory contains an infinite tower of fields. The bi-local propagator can be expressed as a sum over poles, where each term in the sum is a {\em non-standard} AdS$_2$ propagator with non-trivial residues \cite{Jevicki:2016bwu}. Remarkably, this tower can be realized as a Kaluza-Klein tower coming from an additional third dimension \cite{Das:2017pif, Das:2017hrt}. This reproduces the spectrum {\em as well as the propagator}, including the enhanced propagator of \cite{Maldacena:2016hyu}. The kinetic terms are now standard: the nontrivial residues at the poles in the SYK propagator now appear as nontrivial wave functions. However, one should not expect that higher point functions  
\cite{Gross:2017hcz} can be reproduced by local interactions in this 3D picture \cite{Gross:2017aos}.
The usefulness of an additional dimension also appears in the description of Higher Spin theories \cite {Koch:2014aqa}. 

Despite all these successes, the actual emergent space-time of the SYK model (or of any other similar SYK-type models) is not yet understood.
There are several reasons why the  AdS$_2$ or dS$_2$ on which the bi-locals live should {\em not} be considered to be the bulk spacetime in the usual sense of AdS/CFT.  Consider for concreteness the Euclidean partition function. Changing variables to bi-local fields one reaches a solution (the propagator and quadratic fluctuations) which features a Lorentzian signature, coming from the fact that the two points of the bi-local become coordinates of a Lorentizan signature. On the other hand, we expect that the dual theory should live in Euclidean spacetime EAdS$_2$ \cite{Maldacena:2016hyu}. One issue which is detrimental to a potential Lorentzian identification associated with this data comes from the factors of `` $i$ " which inevitably appears in a Lorentzian dual theory, but  absent in the SYK propagator.
Secondly, the radial part of the AdS$_2$ wave functions which appear in the SYK propagator (whether or not we write this in the 3D language) are not the usual normalizable AdS wave functions, but satisfy different boundary conditions. These unusual wave functions are, however, required since these are the ones which diagonalize the SYK kernel \cite{Polchinski:2016xgd, Jevicki:2016bwu}. This suggests that they might be better thought of as dS$_2$ wave functions \cite{Maldacena:2016hyu} \footnote{This has been suggested by J. Maldacena \cite{maldaper}.}.


In this paper we provide a key step towards a resolution of both issues. We will show that a non-local transform relates the bi-local field to a field whose underlying dynamics is in Euclidean AdS$_2$. We will arrive at this transform following the same principles underlying the derivation of the corresponding transform for the $O(N)$ model in $d=3$ \cite{Koch:2010cy, Koch:2014mxa}: the idea is to find a canonical transformation in the four dimensional phase space of the two points in the bi-local such that the symmetries of EAdS$_2$ are realized correctly. This suggets a simple transformation kernel for the momentum space fields. It turns out that the corresponding position space kernel is a $H^2$ Radon transform. Radon transforms have appeared (explicitly or implicitly) in discussions of AdS/CFT, most notably in \cite{Czech:2016xec, deBoer:2016pqk, Bhowmick:2017uci} where this is used to go from the bulk to the kinematic space of the boundary field theory on a time slice.  Indeed the space on which the bi-locals live is a version of kinematic space. However, unlike these papers we are not working on a time slice in the bulk - rather our transform takes unequal Euclidean time fields on EAdS$_2$ to bi-locals. Though mathematically identical, our transform is conceptually somewhat different. The necessity of a Radon transform in this context has been in fact mentioned in \cite{Maldacena:2016hyu}.

This transformation takes the particular combinations of Bessel functions which appear in the SYK propagators to the modified Bessel functions which appear in the standard EAdS$_2$ propagator.In addition we find extra leg factors which  resemble the leg pole factors of the $c=1$ matrix model . In that case these were necessary to relate the collective field \cite{Das:1990kaa} to the tachyon field of the dual 2D string theory and reproduce the $S$-Matrix \cite{Moore:1991ag, legpole} (for a recent improved understanding see \cite{Balthazar:2017mxh}).  The leg poles represent discrete states of the 2D string and analogously it is tempting to suggest that the leg pole factors also arise from similar bulk degrees of freedom, which remain to be identified. An explicit correspondence between the SYK propagator and the propagator of macroscopic loop operators \cite{Moore:1991ag}that we establish supports this interpretation.

The content of the paper is as follows: In section \ref{sec:question}, we review relevant aspects of the bi-local solution of the model and illuminate the dS$_2$ nature of the wave functions. In section \ref{sec:leg transformation}, we introduce the Leg transformations and their meaning. In section \ref{sec:green's function}, Leg factors and the Propagator is discussed. Section \ref{sec:conclusion} is reserved for conclusions.

\section{Question of Dual Spacetime}
\label{sec:question}
In this section, we clarify the question regarding the signature of the SYK dual gravity theory.

The Sachdev-Ye-Kitaev model \cite{Kitaev:2015} is a quantum mechanical many body system with all-to-all interactions on fermionic $N$ sites ($N \gg 1$), described by the Hamiltonian
	\begin{equation}
		H \, = \, \frac{1}{4!} \sum_{i,j,k,l=1}^N J_{ijkl} \, \chi_i \, \chi_j \, \chi_k \, \chi_l \, ,
	\label{eq:Hamiltonian}
	\end{equation}
where $\chi_i$ are Majorana fermions, which satisfy $\{ \chi_i, \chi_j \} = \delta_{ij}$.
The coupling constant $J_{ijkl}$ are random with a Gaussian distribution with width $J$.
The generalization to analogous $q$-point interacting model is straightforward \cite{Kitaev:2015,Maldacena:2016hyu}.
After the disorder averaging for the random coupling $J_{ijkl}$, there is only one effective coupling $J$ in the effective action.
The model is usually treated by replica method.
One does not expect a spin glass state in this model at least in the leading order of $1/N$ \cite{Sachdev:2015efa} so that we can restrict to
the replica diagonal subspace \cite{Jevicki:2016bwu}.
The Large $N$ theory is simply represented through a (replica diagonal) bi-local collective field:
	\begin{equation}
		\Psi(t_1, t_2) \, \equiv \, \frac{1}{N} \sum_{i=1}^N \chi_i(t_1) \chi_i(t_2) \, ,
	\end{equation}
where we have suppressed the replica index. The corresponding path-integral is
	\begin{equation}
		Z \, = \, \int \prod_{t_1, t_2} \mathcal{D}\Psi(t_1, t_2) \ \mu[\Psi] \, e^{-S_{\rm col}[\Psi]} \, , 
	\label{eq:collective partition function}
	\end{equation}
where $S_{\rm col}$ is the collective action:
	\begin{equation}
		S_{\rm col}[\Psi] \, = \, \frac{N}{2} \int dt \, \Big[ \partial_t \Psi(t, t')\Big]_{t' = t} \, + \, \frac{N}{2} \, {\rm Tr} \log \Psi \, - \, \frac{J^2N}{2q} \int dt_1 dt_2 \, \Psi^q(t_1, t_2) \, .
	\label{eq:S_col}
	\end{equation}
Here the trace term comes from a Jacobian factor due to the change of path-integral variable, and the trace is taken over the bi-local time.
One also has an appropriate order $\mathcal{O}(N^0)$ measure $\mu[\Psi]$.
There is another formulation with two bi-local fields: the fundamental fermion propagator $G(t_{12})$ and the self energy $\Sigma(t_{12})$. This is
is equivalent to the above formulation after elimination of $\Sigma(t_{12})$.
In this paper, we focus on this Euclidean time SYK model.

Fluctuations around the critical IR saddle point background $\Psi_0(t_1,t_2)$ can be studied by expanding the bi-local field as \cite{Jevicki:2016bwu}
	\begin{equation}
		\Psi(t_1, t_2) \, = \, \Psi_0(t_1, t_2) \, + \, \frac{1}{\sqrt{N}} \ \overline{\Psi}(t_1, t_2) \, ,
	\end{equation}
where $\Psi_0$ is the IR large $N$ saddle-point solution and $\overline{\Psi}$ is the fluctuation.
At the quadratic level, we have a quadratic kernel $\mathcal{K}$.
The diagonalization of this quadratic kernel is done by the eigenfunction $u_{\nu, \omega}$ and the eigenvalue $\widetilde{g}(\nu)$ as
	\begin{equation}
		\int dt_1' dt_2' \, \mathcal{K}(t_1, t_2; t_1', t_2') \, u_{\nu, \omega}(t_1', t_2') \, = \, \widetilde{g}(\nu) \, u_{\nu, \omega}(t_1, t_2) \, .
	\end{equation}
The quadratic kernel $\mathcal{K}$ is in fact a function of the bi-local $SL(2, \mathcal{R})$ Casimir 
	\begin{align}
		C_{1+2} \, &= \, \big( \hat{D}_1 + \hat{D}_2 \big)^2 \, - \, \frac{1}{2} \big( \hat{P}_1 + \hat{P}_2 \big) \big( \hat{K}_1 + \hat{K}_2 \big)
		\, - \, \frac{1}{2} \big( \hat{K}_1 + \hat{K}_2 \big) \big( \hat{P}_1 + \hat{P}_2 \big) \nonumber\\
		&= \, - \, (t_1 - t_2)^2 \, \partial_1 \partial_2 \, ,
	\label{Casimir}
	\end{align}
with the $SL(2, \mathcal{R})$ generators $\hat{D}=- t \partial_t$, $\hat{P}=\partial_t$, and $\hat{K}=t^2 \partial_t$.
The common eigenfunctions of the bi-local $SL(2, \mathcal{R})$ Casimir (\ref{Casimir}) are, due to the properties of the conformal block, given by the three-point function of the form
	\begin{equation}
		|t_{12}|^{2\Delta} \Big\langle \mathcal{O}_h(t_0) \, \mathcal{O}_{\Delta}(t_1) \, \mathcal{O}_{\Delta}(t_2) \Big\rangle
		\, = \, \frac{{\rm sgn}(t_{12})}{|t_{10}|^h |t_{20}|^h |t_{12}|^{-h} } \, , 
	\end{equation}
where we defined $t_{ij}\equiv t_i-t_j$.
Since the SYK quadratic kernel $\mathcal{K}$ is a function of this bi-local $SL(2, \mathcal{R})$ Casimir, this three-point function is also the eigenfunction of the SYK quadratic kernel.
For the investigation of dual gravity theory, it is more useful to Fourier transform from $t_0$ to $\omega$ by
	\begin{align}
		\Big\langle \widetilde{\mathcal{O}_h}(\omega) \, \mathcal{O}_{\Delta}(t_1) \, \mathcal{O}_{\Delta}(t_2) \Big\rangle
		\, &\equiv \, \int dt_0 \, e^{i \omega t_0} \, \Big\langle \mathcal{O}_h(t_0) \, \mathcal{O}_{\Delta}(t_1) \, \mathcal{O}_{\Delta}(t_2) \Big\rangle \nonumber\\
		&= \, - \, \sqrt{\pi} \, \cot(\pi \nu) \, \Gamma(\tfrac{1}{2}-\nu) \, |\omega|^{\nu} \, \frac{{\rm sgn}(t_{12})}{|t_{12}|^{2\Delta-\frac{1}{2}} } \, e^{i \omega(\frac{t_1+t_2}{2})} 
		Z_{\nu}(|\tfrac{\omega t_{12}}{2}|) \, ,
	\end{align}
where we used $h=\nu+1/2$ and defined
	\begin{equation}
		Z_{\nu}(x) \, = \, J_{\nu}(x) \, + \, \xi_{\nu} \, J_{-\nu}(x) \, , \qquad \xi_{\nu} \, = \, \frac{\tan(\pi \nu/2)+1}{\tan(\pi \nu/2)-1} \, .
	\label{eq:Z_nu}
	\end{equation}
The $t_0$ integral in the Fourier transform can be performed by decomposing the integration region into three pieces.
The complete set of $\nu$ can be understood from the representation theory of the conformal group, as discussed recently in \cite{Kitaev:2017hnr}.
We have the discrete modes $\nu=2n+3/2$ with ($n=0, 1, 2, \cdots$) and the continuous modes $\nu=ir$ with ($0<r<\infty$).
Adjusting the normalization, we define our eigenfunctions by
	\begin{equation}
		u_{\nu, \omega}(t, \hz) \, \equiv \, {\rm sgn}(\hz) \, \hz^{\frac{1}{2}} \, e^{i \omega t} \, Z_{\nu}(|\omega \hz|) \, ,
	\label{eq:eigenfunc}
	\end{equation}
which have normalization condition
	\begin{equation}
		\int_{-\infty}^{\infty} \frac{dt}{2\pi} \int_0^{\infty} \frac{d\hz}{\hz^2} \ u_{\nu, \omega}^*(t, \hz) \, u_{\nu', \omega'}(t, \hz) \, = \, N_{\nu} \, \delta(\nu-\nu') \delta(\omega-\omega') \, ,
	\end{equation} 
with
	\begin{align}
		N_{\nu} \, = \,
		\begin{cases}
			(2\nu)^{-1}  &{\rm for}\ \nu=3/2+2n \\
			2\nu^{-1}\sin\pi\nu \quad &{\rm for}\ \nu=ir \, .
		\end{cases}
	\label{N_nu}
	\end{align}
Here we used the change of the coordinates by 
	\begin{equation}
		t \, \equiv \, \frac{t_1 + t_2}{2} \, , \qquad \hz \, \equiv \, \frac{t_1 - t_2}{2} \, .
	\label{eq:dS_2 coordinates}
	\end{equation}
The bi-local $SL(2, \mathcal{R})$ Casimir can be seen to take the form of a Laplacian of Lorentzian two dimensional 
Anti de-Sitter or de-Sitter space (in this two dimensional  case they are characterized by the same isometry group SO(2,1) or SO(1,2)).
Under the canonical identification with AdS
	\begin{equation}
		ds^2 \, = \, \frac{-dt^2+d\hz^2}{\hz^2} \, ,
	\end{equation}
it equals
	\begin{equation}
		C_{1+2} \, =  \, z^2 (-\partial_t^2 + \partial_{\hz}^2) \, .
	\end{equation}
Consequently the SYK eigenfunctions should be compared  with known AdS$_2$ or dS$_2$ basis wave functions. 

Note that the Bessel function $Z_\nu$ (\ref{eq:Z_nu}) are not the standard normalizable modes used in quantization of scalar fields in AdS$_2$:
in particular they have rather different boundary conditions at the Poincare horizon.
Another important property of this basis is that when viewed as a Schrodinger problem as in \cite{Polchinski:2016xgd} it has a set of bound states, in addition to  the scattering states.
This will be discussed in detail in Section \ref{sec:leg transformation} (see the left picture of FIG. \ref{fig:potentials}).

This leads one to try an identification with de-Sitter basis functions \footnote{This possibility has been emphasized to us by J. Maldacena \cite{maldaper}.}.
As we will see, the bi-local SYK wave functions can be realized as a particular $\alpha$-vacuum of Lorentzian dS$_2$ with a choice of $\alpha=i \pi h = i\pi (\nu+1/2)$.
This is seen as follows.
We consider the dS$_2$ background with a metric given by
	\begin{equation}
		ds^2 \, = \, \frac{-d\eta^2+dt^2}{\eta^2} \, .
	\label{eq:dS-metric}
	\end{equation}
This can be obtained by the coordinate change (\ref{eq:dS_2 coordinates}) by replacing $z\to\eta$.
The Euclidean (Bunch-Davies \cite{Bunch:1978yq}) wave function of a massive scalar field is given by 
	\begin{equation}
		\phi^E_\omega(\eta) \, e^{i\omega t} \, ,
	\end{equation}
with	
	\begin{equation}
		\phi^E_\omega(\eta) \, = \, \eta^{\frac{1}{2}} \, H^{(2)}_{\nu}(|\omega| \eta) \, , \qquad \nu \, = \, \sqrt{\frac{1}{4} -m^2} \, , 
	\end{equation}
where $H^{(2)}_{\nu}$ is the Hankel function of the second kind.
Since the $t$-dependence is always like $e^{i\omega t}$, in the following we will focus only on the $\eta$ dependence.
The $\alpha$-vacuum wave function is defined by Bogoliubov transformation from this Euclidean wave function \cite{Mottola:1984ar, Allen:1985ux} as
	\begin{align}
		\phi^{\alpha}_\omega(\eta) \, &\equiv \, N_{\alpha} \Big[ \phi^E_\omega(\eta) \, + \, e^{\alpha} \phi^{E*}_\omega(\eta) \Big] \nonumber\\
		&= \, N_{\alpha} \, \eta^{\frac{1}{2}} \Big[ H^{(2)}_{\nu}(|\omega| \eta) \, + \, e^{\alpha} H^{(1)}_{\nu}(|\omega| \eta) \Big] \, ,
	\end{align}
where
	\begin{equation}
		N_{\alpha} \, = \, \frac{1}{\sqrt{1-e^{\alpha+\alpha^*}}} \, ,
	\end{equation}
and $\alpha$ is a complex parameter.
Now let us consider a possibility of $\alpha$-vacuum with
	\begin{equation}
		\alpha \, = \, i \pi \left( \nu + \frac{1}{2} \right) \, = \, i \pi h \, .
	\end{equation}
With this choice of $\alpha$, using the definition of the Hankel functions
	\begin{equation}
		H^{(1)}_{\nu}(x) \, = \, \frac{J_{-\nu}(x)-e^{-i\pi \nu} J_{\nu}(x)}{i \sin(\pi \nu)} \, , \qquad 
		H^{(2)}_{\nu}(x) \, = \, \frac{J_{-\nu}(x)-e^{i\pi \nu} J_{\nu}(x)}{-i \sin(\pi \nu)} \, ,
	\end{equation}
one can rewrite the $\alpha$-vacuum wave function as
	\begin{equation}
		\phi^{\alpha}_\omega(\eta) \, = \, \left( \frac{2 \, \eta^{\frac{1}{2}}}{1+\xi_{\nu} \, e^{-i\pi \nu}} \right) \, Z_{\nu}(|\omega| \eta) \, ,
	\end{equation}
where $Z_{\nu}$ is defined in Eq.(\ref{eq:Z_nu}).
After excluding the $\eta$-independent part of the wave function, we can write the $\eta$-dependent part as
	\begin{equation}
		\phi^{\alpha}_\omega(\eta) \, = \, \eta^{\frac{1}{2}} \, Z_{\nu}(|\omega| \eta) \, .
	\end{equation}
This wave function agrees with the eigenfunction of the SYK quadratic kernel (\ref{eq:eigenfunc})
after the identifications of $\eta = (t_1-t_2)/2$ and $t=(t_1+t_2)/2$.

Due to this observation, one might attempt to claim that the dual gravity theory of the SYK model is given by Lorentzian dS$_2$ space. However, there is a critical issue in this claim.
Apart from the Lorentzian signature in this metric (\ref{eq:dS-metric}), we still have a discrepancy in the exponent of the partition function (\ref{eq:collective partition function}) with a factor of ``$i$''.
Namely, if the dual gravity theory (higher spin gravity or string theory) is Lorentzian dS$_2$, it must have
	\begin{equation}
		Z \, = \, \int \mathcal{D}h_n \, \mathcal{D}\Phi_m \, \exp\bigg[ i \Big( S_{{\rm grav}}[h, \Phi] + S_{\rm matter}[h, \Phi] \Big) \bigg] \, , 
	\end{equation}
where we collectively denote the graviton and other ``higher spin'' gauge fields by $h_n$ and the dilaton and other matter fields by $\Phi_m$.
Hence the agreement of the SYK bi-local propagator
	\begin{equation}
		\mathcal{D}_{\rm SYK}(t_1, t_2; t_1', t_2') \, = \, \Big\langle \overline{\Psi}(t_1, t_2) \overline{\Psi}(t_1', t_2') \Big\rangle
		\, = \, \sum_{m=0}^{\infty} G_{p_m}(t_1, t_2; t_1', t_2') \, ,
	\end{equation}
with a dS$_2$ propagator
	\begin{equation}
		\mathcal{D}_{\rm dS}(\eta, t; \eta', t') \, = \, \frac{1}{i} \, \sum_{m=0}^{\infty} \Big\langle \Phi_m(\eta, t) \Phi_m(\eta', t) \Big\rangle
		\, = \, \frac{1}{i} \, \sum_{m=0}^{\infty} G_m(\eta, t; \eta', t') \, ,
	\end{equation}
is only up to the factor $i$.
Namely, even if we have a complete agreement of $G_{p_m}$ with $G_m$ by identifying the coordinates by (\ref{eq:dS_2 coordinates}) (with a replacement of $z \to \eta$),
there is a problem with the signature (i.e. the discrepancy of the factor $i$).
For higher point functions, the same $i$-problem proceeds due to the $i$ factors coming from the propagator and each vertex.

To conclude, for the Euclidean SYK model under consideration, one needs a dual gravity theory to be in the hyperbolic plane H$_2$ (i.e. Euclidean AdS$_2$)
for the matching of $n$-point functions.
We will set the basis for the EAdS$_2$ realization in the next section.

\section{Transformations and Leg Factors}
\label{sec:leg transformation}
As we have commented in the Introduction in order to identify an Euclidean bulk dual description (rather than a Lorentzian),
we will need a transformation which brings the SYK eigenfunctions (as given on bi-local space-time) to the standard eigenfunctions of the EAdS$_2$ Laplacian.
We will arrive at this transformation by considering the bi-local map described in \cite{Koch:2010cy, Koch:2014aqa} for higher dimensional case.
In our current $d=1$ case, the map is even simpler. It will be seen to take the form of a $H^2$ Radon transform (a related suggestion was made in \cite{Maldacena:2016hyu}).
The need for a non-local transform on external legs appears to be characteristic of collective theory (which as a rule contains a minimal set of physical degrees of freedom).
The first appearance of Radon type transforms in identifying holographic space-time was seen in the $c=1$ / $D=2$ string correspondence.
\footnote{
The transformation introduced in \cite{Jevicki:1993zg} from the collective to a 2D (black hole) space-time took the form  
	\begin{equation}
		T(u, v) \, = \, \int_{-\infty}^{\infty} dt \int_0^{\infty} dx \, \delta \left( \frac{u e^{-t} + v e^{t}}{2} - x^2 \right) \, \gamma(i \partial_t) \, \phi(t, x) \, ,
	\end{equation}
where $T(u, v)$ is the tachyon field in the Kruskal coordinates representing the target space-time and $\phi(t, x)$ is related to the eigenvalue density field.
Related maps from the collective field or fermions to fields in a black hole background have been proposed in
\cite{Das:1992dw} which are also possibly related to Radon transforms.}
This is seen precisely in the form of what is known as the regular Radon transform.

Let us describe procedure formulated  in \cite{Koch:2010cy, Koch:2014aqa} for constructing the bi-local to space-time map.
The method is based on construction of  canonical transformations in phase space : bi-local ($t_1, p_1$), ($t_2, p_2$) and EAdS$_2$ ($\tau, p_\tau$), ($z, p_z$).
We consider the Poincare coordinates for the Euclidean AdS$_2$ space-time 
	\begin{equation}
		ds^2 \, = \, \frac{d\tau^2+dz^2}{z^2} \, .
	\label{eq:EAdS metric}
	\end{equation}
One way to obtain the bi-local map is to equate the $SL(2, \mathcal{R})$ generators.
	\begin{equation}
		\hat{J}_{1+2} \, = \, \hat{J}_{\rm EAdS} \, .
	\end{equation}
The one-dimensional bi-local conformal generators are
	\begin{equation}
		\hat{D}_{1+2} \, = \, t_1 \, p_1 + t_2 \, p_2 \, , \qquad \hat{P}_{1+2} \, = \, - p_1 - p_2 \, , \qquad
		\hat{K}_{1+2} \, = \, - \, t_1^2 \, p_1 \, - \, t_2^2 \, p_2 \, ,
	\end{equation}
and the EAdS$_2$ generators are given by 
	\begin{equation}
		\hat{D}_{\rm EAdS} \, = \, \tau \, p_{\tau} + z \, p_z \, , \qquad \hat{P}_{\rm EAdS} \, = \, - p_{\tau} \, , \qquad
		\hat{K}_{\rm EAdS} \, = \, (z^2 - \tau^2) \, p_{\tau} \, - \, 2 \tau z \, p_z \, ,
	\end{equation}
where we defined $p_1\equiv-\partial_{t_1}$, $p_2\equiv-\partial_{t_2}$, $p_{\tau}\equiv-\partial_{\tau}$, $p_{z}\equiv-\partial_z$.
Equating the generators, we can determine the map.
From the $\hat{P}$ generators, we have $p_{\tau} = p_1 + p_2$. Using this result for the other generators, we get two equations to solve:
	\begin{align}
		z \, p_z \, &= \, (t_1 - \tau) p_1 \, + \, (t_2 - \tau) p_2 \nonumber\\
		- \, z^2 \, p_\tau \, &= \, (t_1 - \tau)^2 p_1 \, + \, (t_2 - \tau)^2 p_2 \, .
	\end{align}
These are solved by
	\begin{equation}
		\tau \, = \, \frac{t_1 \, p_1 - t_2 \, p_2}{p_1-p_2} \, , \ \quad p_{\tau} \, = \, p_1 + p_2 \, , \ \quad
		z^2 \, = \, - \left( \frac{t_1 - t_2}{p_1 - p_2} \right)^2 p_1 p_2 \, , \ \quad p_z^2 \, = \, - 4 p_1 p_2 \, .
	\end{equation}
One can see that the canonical commutators are preserved under the transform (at least classically, i.e. in terms of the Poisson bracket).
Namely, $[\tau, p_{\tau}]=[z, p_z]=1$ and others vanish provided that $[t_i, p_j]=\delta_{ij}$, with ($i,j=1,2$).
Hence, we conclude the map is canonical transformation, which is also a point transformation in momentum space.
For the kernel which implements this momentum space  correspondence we can take 
	\begin{equation}
		\mathcal{R}(p_1, p_2; p_{\tau}, p_z) \, = \, \frac{\delta(p_{\tau}- (p_1+p_2))}{\sqrt{p_z^2 + 4p_1p_2}} \, .
	\end{equation}
Through Fourier transforming all momenta to corresponding coordinates, the associated coordinate space kernel becomes
\footnote{Here, we have ignored possible issues related to the range of variables.} 
	\begin{equation}
		\mathcal{R}(t_1, t_2; \tau, z) \, =  \, \delta(\eta^2 - (\tau-t)^2 -z^2) \, .
	\end{equation}
	With an additional multiplicative factor of of $\eta$ this is known as the Circular Radon transform (\ref{eq:geodesic Radon transf}) which has a simple relationship to Radon transform on $H^2$.

There is another construction of the Radon transform which is used in \cite {Czech:2016xec, deBoer:2016pqk, Bhowmick:2017uci} and is based on integration over geodesics.
For the Euclidean AdS$_2$ space-time (\ref{eq:EAdS metric}), a geodesic is given by a semicircle
	\begin{equation}
		(\tau - \tau_0)^2 \, + \, z^2 \, = \, \frac{1}{E^2} \, ,
	\label{eq:geodesic}
	\end{equation}
where $\tau=\tau_0$ is the center of the semicircle and $1/E$ is the radius.
The Radon transform of a function of the bulk coordinates $f(\tau, z)$ is a function of the parameters of a geodesic ($E, \tau_0$) defined by
	\begin{equation}
		\big[ \mathcal{R} f\big](E, \tau_0) \, \equiv \, \int_{\gamma} ds \, f(\tau, z(\tau)) \, ,
	\end{equation}
where the integral is over the geodesic.
From the geodesic equation (\ref{eq:geodesic}), this transform is explicitly written as
	\begin{equation}
		\big[\mathcal{R} f\big](\eta, t)
		\, = \, 2\eta \int^{t+\eta}_{t-\eta} d\tau \int_0^{\infty} \frac{dz}{z} \, \delta\Big( \eta^2 - (\tau - t)^2 -z^2 \Big) \, f\left(\tau, z \right) \, ,
	\label{eq:geodesic Radon transf}
	\end{equation}
where we have used the identifications $1/E=\eta$ and $\tau_0=t$; the resulting function $[\mathcal{R} f](\eta, t)$ is understood as a function on the Lorentzian dS$_2$ (\ref{eq:dS-metric}).

We will now explicitly evaluate the Radon transformation of (unit-normalized) EAdS$_2$ wave functions (see Appendix \ref{app:unit normalized eads/ds wave functions})
	\begin{equation}
		\overline{\phi}_{\rm EAdS_2}(\tau, z) \, = \, \alpha_{\nu} \, z^{\frac{1}{2}} \, e^{-i\omega \tau} \, K_{\nu}(|\omega|z) \,
	\end{equation}
From the above formula of the Radon transform (\ref{eq:geodesic Radon transf}), we get 
	\begin{equation}
		\Big[ \mathcal{R} \overline{\phi}_{\rm EAdS_2} \Big](\eta, t)
		\, = \, \alpha_{\nu} \, \eta \int_{t-\eta}^{t+\eta} \frac{d\tau}{\eta^2-(\tau-t)^2} \ (\eta^2-(\tau-t)^2)^{\frac{1}{4}} \, e^{-i\omega \tau} \, K_{\nu}(|\omega|\sqrt{\eta^2-(\tau-t)^2}) \, .
	\end{equation}
Now shifting the integral variable $\tau \to \tau+t$ and using the symmetry of the integrand, one can rewrite this integral as
	\begin{equation}
		\Big[ \mathcal{R} \overline{\phi}_{\rm EAdS_2} \Big](\eta, t)
		\, = \, 2 \, \alpha_{\nu} \, \eta \, e^{-i\omega t} \int_0^{\eta} d\tau \, \left( \frac{1}{\eta^2-\tau^2} \right)^{\frac{3}{4}} \cos(\omega \tau) \, K_{\nu}(|\omega|\sqrt{\eta^2-\tau^2}) \, .
	\end{equation}
Further rewriting the $\cos(\omega \tau)$ in terms of $J_{-1/2}(\omega \tau)$ and changing the integration variable to $\tau=\eta \sin \theta$, we find
	\begin{align}
		\Big[ \mathcal{R} \overline{\phi}_{\rm EAdS_2} \Big](\eta, t) \, &= \, \sqrt{\frac{\pi^3}{2}} \, \frac{\alpha_{\nu} |\omega|^{\frac{1}{2}} \eta}{\sin(\pi \nu)} \, e^{-i\omega t} \nonumber\\
		&\ \times \int_0^{\frac{\pi}{2}} d\theta \, (\tan\theta)^{\frac{1}{2}} \, J_{-\frac{1}{2}}(|\omega|\eta \sin \theta)
		\, \Big[ I_{-\nu}(|\omega|\eta \cos\theta) \, - \, I_{\nu}(|\omega|\eta \cos\theta) \Big] \, ,
	\end{align}
where we decomposed the modified Bessel function of the second kind into two first kinds.
This $\theta$ integral is indeed given in Eq.(4) of $12\cdot11$ of \cite{Watson}, which leads to
	\begin{equation}
		\Big[ \mathcal{R} \overline{\phi}_{\rm EAdS_2} \Big](\eta, t)
		\, = \, - 2i\sqrt{\pi} \, \frac{\Gamma(\frac{1}{4}+\frac{\nu}{2})}{\Gamma(\frac{3}{4}+\frac{\nu}{2})} \, \beta_{\nu} \, \eta^{\frac{1}{2}} \, e^{-i\omega t}
		\Bigg[ J_{\nu}(|\omega| \eta) \, + \, \frac{\tan\frac{\pi \nu}{2}+1}{\tan\frac{\pi \nu}{2}-1} \, J_{-\nu}(|\omega| \eta) \Bigg] \, ,
\label{radon1}
	\end{equation}
where we also used Eq.(\ref{alpha-beta}).
The inside of the square bracket precisely agrees with the particular combination of Bessel functions, $Z_{\nu}(|\omega| \eta)$ function defined in Eq.(\ref{eq:Z_nu}).

When $\nu_n = 3/2 + 2n$ the second term in this square bracket vanishes.
As will be clear soon, we need the radon transform of the modified Bessel function $I_{\nu_n}$ with. This can be likewise evaluated to yield
\ben
\mathcal{R} [\alpha^\prime_{\nu_n} z^{1/2} e^{-ik\tau} I_{\nu_n} (|k|z) ] = (2\nu_n \eta)^{1/2} e^{-ikx} J_{\nu_n} (|k|\eta)
\label{disc-radon}
\een
where
\ben
\alpha^\prime_{\nu_n} = \left( \frac{2\nu_n}{\pi} \right)^{\frac{1}{2}} 
\frac{\Gamma (\frac{3}{4}+\frac{\nu_n}{2})}{\Gamma(\frac{1}{4}+\frac{\nu_n}{2})}
\label{alphaprime}
\een
The extra $\nu$-dependent factor in (\ref{radon1}) which appears in front of the unit-normalized dS$_2$ wave function described in Appendix \ref{app:unit normalized eads/ds wave functions}
should be understood as a leg factor (\ref{eq:leg factor}). 
As we will see later,
this is analogous to what happens in the $c=1$ matrix model \cite{Moore:1991ag, legpole}.

In summary, we have the Radon transform
	\begin{equation}
		\mathcal{R} \, \overline{\phi}^{(\rm EAdS_2)}_{\omega, \nu}(\tau, z) \, = \, L(\nu) \, \overline{\psi}^{(\rm dS_2)}_{\omega, \nu}(\eta, t) \, ,
	\label{eq:Radon transf}
	\end{equation}
where $\overline{\phi}_{\rm EAdS_2}$ and $\overline{\psi}_{\rm dS_2}$ are the unit-normlized wave functions defined in Eq.(\ref{phi_EAdS}) and Eq.(\ref{psi_dS}), respectively,
while the leg factor is defined by
	\begin{equation}
		L(\nu) \, \equiv \, ({\rm Leg\ Factor}) \, = \, - 2i\sqrt{\pi} \, \frac{\Gamma(\frac{1}{4}+\frac{\nu}{2})}{\Gamma(\frac{3}{4}+\frac{\nu}{2})} \, .
	\label{eq:leg factor}
	\end{equation}

The inverse transformations are
	\begin{equation}
		\mathcal{R}^{-1} \, \overline{\psi}^{(\rm dS_2)}_{\omega, \nu}(\eta, t) \, = \, L^{-1}(\nu) \, \overline{\phi}^{(\rm EAdS_2)}_{\omega, \nu}(\tau, z) \, .
	\label{eq:inverse transf}
	\end{equation}
for $\nu \neq 3/2+2n$, while for $\nu = 3/2+2n$ we have instead
\ben
\mathcal{R}^{-1} \, \overline{\psi}^{(\rm dS_2)}_{\omega, \nu_n}(\eta, t) \, = \,
\alpha^\prime_{\nu_n} z^{1/2} e^{-ik\tau} I_{\nu_n} (|k| z)
\label{invraddisc}
\een

\begin{figure}[t!]
	\vspace{-10pt}
	\begin{center}
		\scalebox{0.9}{\hspace{-10pt}\includegraphics{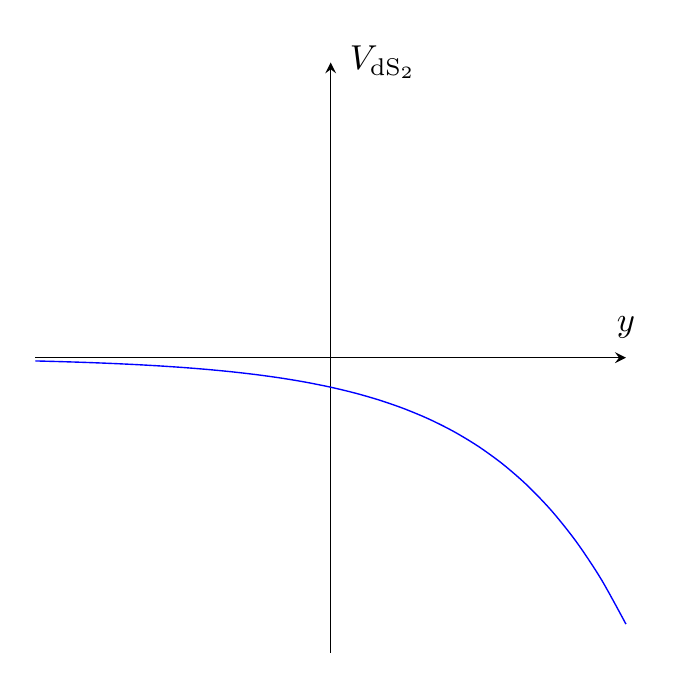} \qquad \includegraphics{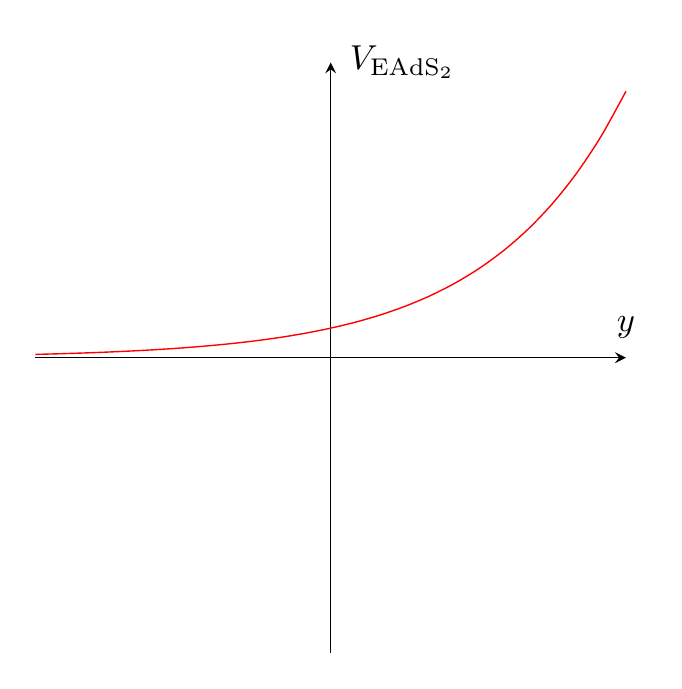}}
	\end{center}
	\vspace{-10pt}
	\caption{The de Sitter potential $V_{{\rm dS}_2}$ has bound states and scattering states. On the other hand, the Euclidean AdS potential $V_{{\rm AdS}_2}$ has only scattering modes.}
	\label{fig:potentials}
\end{figure}

Under the Radon transform $\mathcal{R}$, the Laplacian of Lorentzian dS$_2$ is transformed into that of Euclidean AdS$_2$:
	\begin{equation}
		\Box_{{\rm ds}_2} \, \psi_{{\rm dS}_2}(\eta, t) \, = \, - \, \mathcal{R} \, \Box_{{\rm EAdS}_2} \, \phi_{{\rm EAdS}_2}(\tau, z) \, ,
	\label{eq:Laplacian Radon}
	\end{equation}
with
	\begin{equation}
		\Box_{{\rm ds}_2} = \, \eta^2 (- \partial_{\eta}^2+\partial_t^2) \, , \qquad \Box_{{\rm EAdS}_2} = \, z^2 (\partial_{\tau}^2 + \partial_z^2) \, .
	\label{Laplacian}
	\end{equation}
Here, $\psi_{{\rm dS}_2} = \mathcal{R}\phi_{{\rm EAdS}_2}$ . This role of the Radon transform was first suggested in\ cite{Balasubramanian:2002zh}.

In the rest of this section, we will show that the Radon transformation flips the sign of the potential appearing in the equivalent Schrodinger problem as formulated in 
\cite{Polchinski:2016xgd}.
We start from the Radon transformation (\ref{eq:Laplacian Radon}).
Expanding the wave functions by 
	\begin{align}
		\psi_{{\rm dS}_2}(\eta, t) \, &= \, \eta^{\frac{1}{2}} \sum_\omega e^{-i\omega t} \, \widetilde{\psi}_{{\rm dS}_2}(\eta; k) \, , \nonumber\\
		\phi_{{\rm EAdS}_2}(\tau, z) \, &= \, z^{\frac{1}{2}} \sum_\omega e^{-i\omega \tau} \, \widetilde{\phi}_{{\rm EAdS}_2}(\omega; z) \, , 
	\end{align}
we have corresponding Bessel equations for $\widetilde{\psi}_{{\rm dS}_2}$ and $\widetilde{\phi}_{{\rm EAdS}_2}$.
By changing the coordinates by $y\equiv \log(\omega \eta)$ or $y\equiv \log(\omega z)$, these Bessel equations are reduced to the Schrodinger equations as
	\begin{align}
		\Big( - \partial_y^2 - e^y \Big) \, \widetilde{\psi}_{{\rm dS}_2} \, &= \, - \nu^2 \, \widetilde{\psi}_{{\rm dS}_2} \, , \nonumber\\
		\Big( - \partial_y^2 + e^y \Big) \, \widetilde{\phi}_{{\rm EAdS}_2} \, &= \, - \nu^2 \, \widetilde{\phi}_{{\rm EAdS}_2} \, .
	\end{align}
Therefore, the Radon transform flips the sign of the corresponding Schrodinger potential (see FIG. \ref{fig:potentials}).
The de Sitter potential $V_{{\rm dS}_2} = - e^y$ has bound states as well as scattering states.
On the other hand, the Euclidean AdS potential $V_{{\rm AdS}_2} = e^y$ has only scattering modes. The difference is accounted by the leg pole factors.

\section{Green's Functions and Leg Factors}
\label{sec:green's function}
In this section, we start from the SYK bi-local propagator \cite{Jevicki:2016bwu, Das:2017pif}.
Applying  the inverse Radon transformation (\ref{eq:inverse transf}),
we will show that the resulting propagator can be written in terms of  $EAdS_2$ wave-functions additional momentum space  leg-factors. 

The SYK bi-local propagator is given by 
	\begin{align}
		G(t_1, t_2; t_1', t_2') \, \propto \, J^{-1} \, \int_{-\infty}^{\infty} d\omega \, \sum_{\nu} \,
		\frac{u_{\nu, \omega}^*(t_1, t_2) u_{\nu, \omega}(t_1', t_2')}{N_{\nu} [\widetilde{g}(\nu) - 1]} \, ,
	\label{SYK-propagator}
	\end{align}
where $u_{\nu, \omega}$ are the eigenfunctions defined in Eq.(\ref{eq:eigenfunc}).
Here the summation over $\nu$ is a short-hand notation denotes the discrete mode sum and the continuous mode sum. with the identification  $\eta = (t_1-t_2)/2$ and $t=(t_1+t_2)/2$ the propagator is written in terms of the dS$_2$ wave functions as
	\begin{align}
		G(\eta, t; \eta', t') \, &= \, 2\pi J^{-1} \, \int_{-\infty}^{\infty} d\omega
		\Bigg\{ \sum_{n=0}^{\infty} \, \frac{4\sin\pi \nu_n}{\widetilde{g}(\nu_n) - 1} \ \overline{\psi}_{\omega, \nu_n}^{\, *}(\eta, t) \, \overline{\psi}_{\omega, \nu_n}(\eta', t') \nonumber\\
		&\hspace{100pt} + \, \int_0^{\infty} dr \ \frac{\overline{\psi}_{\omega, \nu}^{\, *}(\eta, t) \, \overline{\psi}_{\omega, \nu}(\eta', t')}{\widetilde{g}(\nu) - 1} \bigg|_{\nu=i r} \Bigg\} \, ,
	\label{eq:dS-propagator}
	\end{align}
where $\nu_n=2n+3/2$.
Next, we use the inverse Radon transform (\ref{eq:inverse transf}) to bring the dS wave functions into the EAdS wave functions.
	\begin{align}
		G(\tau, z; \tau', z') \, &= \, 2\pi J^{-1} \, \int_{-\infty}^{\infty} d\omega \Bigg\{ \sum_{n=0}^{\infty} \, \frac{4\sin\pi \nu_n}{\widetilde{g}(\nu_n) - 1} \
		|L^{-1}(\nu_n)|^2 \ \overline{\phi}_{\omega, \nu_n}^{\, *}(\tau, z) \, \overline{\phi}_{\omega, \nu_n}(\tau', z') \nonumber\\
		&\hspace{100pt} + \, \int_0^{\infty} dr \ |L^{-1}(\nu)|^2 \
		\frac{\overline{\phi}_{\omega, \nu}^{\, *}(\tau, z) \, \overline{\phi}_{\omega, \nu}(\tau', z')}{\widetilde{g}(\nu) - 1} \bigg|_{\nu=i r} \Bigg\} \, .
	\label{eq:EAdS-propagator}
	\end{align}
Here we have denoted  $\overline{\phi}_{\omega, \nu_n}(\tau, z)$ as
\ben
\overline{\phi}_{\omega, \nu_n}(\tau, z) = \alpha^\prime_{\nu_n} z^{1/2} e^{-ik\tau} I_{\nu_n} (|k| z)
\label{defdsdisc}
\een

We will directly evaluate the continuous mode summation for the full Green's function with the  leg factor contribution  in the integrand.
For clarity let us first formally feature the leg factors as Bessel differential operators,as
	\begin{align}
		G(\tau, z; \tau', z') \, &= \, 2\pi J^{-1} \, \big| L^{-1}(\hat{p}_{{\rm EAdS}_2}) \big|^2 \, \int_{-\infty}^{\infty} d\omega \Bigg\{ \sum_{n=0}^{\infty} \,
		\frac{4\sin\pi \nu_n}{\widetilde{g}(\nu_n) - 1} \ \overline{\phi}_{\omega, \nu_n}^{\, *}(\tau, z) \, \overline{\phi}_{\omega, \nu_n}(\tau', z') \nonumber\\
		&\hspace{160pt} + \, \int_0^{\infty} dr \ \frac{\overline{\phi}_{\omega, \nu}^{\, *}(\tau, z) \, \overline{\phi}_{\omega, \nu}(\tau', z')}{\widetilde{g}(\nu) - 1} \bigg|_{\nu=i r} \Bigg\} \, .
	\end{align}
with 
	\begin{equation}
		\hat{p}_{{\rm EAdS}_2} \, \equiv \, \sqrt{\Box_{{\rm EAdS}_2} + \frac{1}{4} \, } \, ,
	\end{equation}
where the Laplacian of $ EAdS_2$ is defined in Eq.(\ref{Laplacian}) and the factors are now acting on standard propagators $ EAdS_2$ .
The above expression for the leg factor differential operators is slightly ambiguous. What we mean is that one of the leg factor differential operator is acting on ($\tau, z$) and the other leg factor operator is acting on ($\tau', z'$).

We now proceed with our off-shell expression of the propagator (\ref{eq:EAdS-propagator})
and evaluation of  the continuous mode summation for the Green's function with  leg factors in the integrand:
	\begin{equation}
		\mathcal{I}_{\rm cont} \, \equiv \, 
		\int_0^{\infty} dr \ |L^{-1}(\nu)|^2 \ \frac{\overline{\phi}_{\omega, \nu}^{\, *}(\tau, z) \, \overline{\phi}_{\omega, \nu}(\tau', z')}{\widetilde{g}(\nu) - 1} \bigg|_{\nu=i r} \, .
	\end{equation}
We evaluate this integral as a contour integral as before.
We note that since the modified Bessel function $K_{\nu}$ is regular on the entire $\nu$-complex plane, we have two sets of poles:
(i). $\nu=p_m$, with $(m=0, 1, 2, \cdots)$.
(ii). $\nu=\nu_n=2n+3/2$, with $(n=0, 1, 2, \cdots)$ where $\Gamma(\frac{3}{4}-\frac{\nu}{2})=\infty$.
After evaluating the residues at these poles, we find the integral as
	\begin{align}
		\mathcal{I}_{\rm cont} \, &= \, \frac{|zz'|^{\frac{1}{2}}}{4\pi^2} \, e^{-i\omega(\tau-\tau')} \, \Bigg\{ \sum_{m=0}^{\infty} \,
		\frac{\Gamma(\frac{3}{4}+\frac{p_m}{2})\Gamma(\frac{3}{4}-\frac{p_m}{2})}{\Gamma(\frac{1}{4}+\frac{p_m}{2})\Gamma(\frac{1}{4}-\frac{p_m}{2})} \,
		\frac{p_m}{\widetilde{g}'(p_m)} \, K_{p_m}(|\omega| z^>) I_{p_m}(|\omega| z^<) \nonumber\\
		&\hspace{95pt} + \, \frac{2}{\pi} \sum_{n=0}^{\infty} \frac{\Gamma^2(\frac{3}{4}+\frac{\nu_n}{2})}{\Gamma^2(\frac{1}{4}+\frac{\nu_n}{2})} 
		\left( \frac{\nu_n}{\widetilde{g}(\nu_n) -1} \right) \, K_{\nu_n}(|\omega| z^>) I_{\nu_n}(|\omega| z^<) \Bigg\} \, .
	\end{align}
The second line in the RHS looks similar to the discrete mode contribution to the propagator (\ref{eq:EAdS-propagator}).
However, these two contributions do not cancel each other. Hence there are two types of the contributions to the final result as
	\begin{align}
		&\quad G(\tau, z; \tau', z') \nonumber\\
		&= \, \frac{|zz'|^{\frac{1}{2}}}{2\pi J} \, \int_{-\infty}^{\infty} d\omega \, e^{-i\omega(\tau-\tau')} \,
		\Bigg\{ \sum_{m=0}^{\infty} \, \frac{\Gamma(\frac{3}{4}+\frac{p_m}{2})\Gamma(\frac{3}{4}-\frac{p_m}{2})}
		 {\Gamma(\frac{1}{4}+\frac{p_m}{2})\Gamma(\frac{1}{4}-\frac{p_m}{2})} \, \frac{p_m}{\widetilde{g}'(p_m)} \, K_{p_m}(|\omega| z^>) I_{p_m}(|\omega| z^<) \nonumber\\
		&\hspace{32pt} +  \, \sum_{n=0}^{\infty} \, \frac{\Gamma^2(\frac{3}{4}+\frac{\nu_n}{2})}{\Gamma^2(\frac{1}{4}+\frac{\nu_n}{2})} 
		\left( \frac{\nu_n}{\widetilde{g}(\nu_n) -1} \right) \, I_{\nu_n}(|\omega| z^<) \Big[ 2I_{\nu_n}(|\omega| z^>) - I_{-\nu_n}(|\omega| z^>) \Big] \Bigg\} \, .
	\end{align}
Of course, here we still have the zero mode ($p_0=3/2$) problem coming from $\Gamma(\frac{3}{4}-\frac{p_0}{2})=\infty$.
In this expression, the Bessel function part of the first contribution in the RHS is the standard form for EAdS propagator,
while the extra factor coming from the leg-factors can be possibly understood as a contribution from the naively pure gauge degrees of freedom as in the $c=1$ model (c.f. \cite{legpole}), in which case
the second contribution in RHS represents the contribution from these modes as in \cite{Moore:1991ag}.

In \cite{Das:2017pif, Das:2017hrt}, we presented at 3D picture of the SYK theory, based on the fact that the non-trivial spectrum predicted by the model,
which are solutions of $\widetilde{g}(p_m)=1$ with $(m=0, 1, 2, \cdots)$ can be reproduced through Kaluza-Klein mechanism  in one higher dimension. This picture is more natural in the $AdS_2$ interpretation of the bilocal space
Now, we will point out a similarity between the 3D picture of the SYK model \cite{Das:2017pif, Das:2017hrt} and the $c=1$ Liouville theory (2D string theory)
\cite{Das:1990kaa, Moore:1991ag, legpole}.

In the 3D description we have a scalar field $\Phi$ 
	\begin{equation}
		S_{\rm 3D} \, = \, \frac{1}{2} \int dx^3 \sqrt{-g} \Big[ - g^{\mu\nu} \partial_{\mu} \Phi \partial_{\nu} \Phi - m_0^2 \Phi^2 - V(y) \Phi^2 \Big] \, ,
	\end{equation}
with a background metric
	\begin{equation}
		ds^2 \, = \, \frac{-dt^2 + d\hz^2}{\hz^2} \, + \, \left( 1 + \frac{a}{\hz} \right)^2 dy^2 \, ,
	\label{eq:3D metric}
	\end{equation}
where $a \sim J^{-1}$, but here we only consider the leading in $1/J$ contribution and suppress the subleading contributions coming from the $yy$-component of the metric.
The detail of the potential $V(y)$ depends on $q$ and for that readers should refer to \cite{Das:2017pif, Das:2017hrt}.
The propagator for the scalar field in this background in the leading order of $1/J$ is given by
	\begin{equation}
		G^{(0)}(\hz,t,y; \hz',t',y') \, = \, |\hz\hz'|^{\frac{1}{2}} \sum_{k} f_{k}(y) f_{k}(y') \int \frac{d\omega}{2\pi} \, e^{-i\omega(t-t')}
		\int \frac{d\nu}{N_{\nu}} \, \frac{Z^*_{\nu}(|\omega \hz|) \, Z_{\nu}(|\omega \hz'|)}{\nu^2 - k^2} \, ,
	\end{equation}
where $f_k(y)$ is the wave function along the third direction $y$ with momentum $k$.
This is simply a rewriting the propagator (\ref{eq:dS-propagator}) by treating the non-local kernel (eigenvalue) by an extra dimension.
The identical procedure leads to the leg-factors.
After the (inverse) Radon transform and the contour integral for the continuous mode sum, the propagator is reduced to 
	\begin{align}
		&\quad G^{(0)}_{\omega;-\omega}(z,y; z',y') \nonumber\\
		&= \, \frac{|z z'|^{\frac{1}{2}}}{4\pi} \sum_{k} f_k(y) f_{k}(y') \,
		\Bigg\{ \, \frac{\Gamma(\frac{3}{4}+\frac{k}{2})\Gamma(\frac{3}{4}-\frac{k}{2})}
		 {\Gamma(\frac{1}{4}+\frac{k}{2})\Gamma(\frac{1}{4}-\frac{k}{2})} \, K_k(|\omega| z^>) I_k(|\omega| z^<) \nonumber\\
		&\hspace{40pt} + \, 2 \sum_{n=0}^{\infty} \, \frac{\Gamma^2(\frac{3}{4}+\frac{\nu_n}{2})}{\Gamma^2(\frac{1}{4}+\frac{\nu_n}{2})} 
		\left( \frac{\nu_n}{\nu_n^2 -k^2} \right) \, I_{\nu_n}(|\omega| z^<) \Big[ 2I_{\nu_n}(|\omega| z^>) - I_{-\nu_n}(|\omega| z^>) \Big] \Bigg\} \, .
	\end{align}

On the other hand, for the $c=1$ matrix model / 2D string duality, the Wilson loop operator is related to the matrix eigenvalue density field $\phi$ by 
	\begin{equation}
		W(t, \ell) \, \equiv \, {\rm Tr}\Big( e^{-\ell M(t)} \Big) \, = \, \int_0^{\infty} dx \, e^{-\ell x} \, \phi(t, x) \, .
	\label{eq:Wilson loop}
	\end{equation}
The corresponding propagator was found by Moore and Seiberg \cite{Moore:1991ag} as
	\begin{align}
		\Big\langle w(t, \varphi) w(t', \varphi') \Big\rangle \, = \, \int_{-\infty}^{\infty} dE \int_0^{\infty} dp \ \frac{p}{\sinh\pi p} \, \frac{\phi^*_{E, p}(t, \varphi) \phi_{E, p}(t', \varphi')}{E^2 - p^2} \, ,
	\label{eq:W-propagator}
	\end{align}
with $\ell=e^{-\varphi}$ and the normalized wave function
	\begin{align}
		\phi_{E, p}(t, \varphi) \, = \, \sqrt{p \sinh\pi p} \, e^{-iEt} \, K_{ip}(\sqrt{\mu} e^{-\varphi}) \, .
	\label{c=1 wave func}
	\end{align}
After evaluating the $p$-integral as a contour integral, we obtain the propagator as
	\begin{align}
		\Big\langle w(t, \varphi) w(t', \varphi') \Big\rangle \, &= \, - \pi \int_{-\infty}^{\infty} dE \, e^{-i E(t-t')} 
		\Bigg\{ \frac{\pi E}{2\sinh \pi E} \, K_{iE}(\sqrt{\mu} e^{-\varphi^<}) \, I_{iE}(\sqrt{\mu} e^{-\varphi^>}) \nonumber\\
		&\hspace{100pt} + \, \sum_{n=1}^{\infty} \, \frac{(-1)^n n^2}{E^2+n^2} \, K_n(\sqrt{\mu} e^{-\varphi^<}) \, I_n(\sqrt{\mu} e^{-\varphi^>}) \Bigg\} \, .
	\label{eq:W-propagator2}
	\end{align}

The point we want to make here is that this 3D picture is completely parallel to the $c=1$ Liouville theory (2D string theory)
\cite{Das:1990kaa, Moore:1991ag, legpole}.
Namely, if we make a change of coordinate by $z=e^{-\varphi}$, then the $\varphi$-direction becomes the Liouville direction,
while the $y$-direction (at least in the leading order of $1/J$) can be understood as the $c=1$ matter direction.
In this comparison, the $\tau$-direction serves as an extra direction.
Finally, the $\nu$ appearing in the SYK model is realized as a momentum $k$ along the $y$-direction in the 3D picture (\ref{eq:3D metric}).
Therefore, we have the following correspondence between the $c=1$ Liouville theory and the 3D picture of the SYK model.

\begin{center}
	\begin{tabular}{ |c|c| }
	\qquad $c=1$ \quad \, \, & \qquad 3D SYK \quad \, \\ \hline
		$ie^{-\varphi}$ & $z$ \\ 
		$-it$ & $y$ \\ 
		$ip$ & $\nu$ \\ 
		$iE$ & $k$ \\ 
		$\sqrt{\mu}$ & $|\omega|$ \\ 
	\end{tabular}
\end{center}

\section{Conclusion}
\label{sec:conclusion}
We have in the present work addressed the question of what represents the bulk dual space-time in the
Sachdev-Ye-Kitaev model. At the outset the question seems simple since  the small fluctuations of the
(Euclidean) SYK model are completely given by a set of Lorentzian wave functions associated with the 
$SL(2,\mathcal{R})$ isometry group. With a simple identification of space-time 
	these are seen to associated with eigenfunctions in de-Sitter (or Anti de-Sitter) space-time (as was discussed in \cite{Kitaev:2015, Polchinski:2016xgd, Jevicki:2016bwu, Maldacena:2016hyu}). And as we have noted these  wave functions are  in correspondence with a particular $\alpha$-vacuum wavefunctions of dS$_2$ space-time. Likewise the propagator and higher point $n$-functions continue to feature this Lorentzian space-time structure.

Even though the Lorentzian bulk dual interpretation seems to be straightforwardly associated with the SYK bi-local data, we have stressed that there is a problem with this interpretation. In most naive sense one would essentially expect that an Euclidean CFT should lead to an Euclidean bulk dual. 
In the case of the SYK model there is a caveat since a role is played by random tensor couplings, whose bulk interpretation is even more unclear.
However, concentrating on the effective bi-local Large $N$ version of the theory we have in this paper provided a resolution, which follows from a further nonlocal redefinition of space-time.
This comes in terms of Leg transformations of Green's functions which place the theory in Euclidean AdS dual setting. Such transformations are actually characteristic of collective field representations of Large $N$ theories.
The leg transformations that we explicitly implement (apart from providing the EAdS$_2$ space-time setting) also bring out the couplings of additional ``discrete'' states. Since this is implemented on all n-point functions it represents a highly nonlinear effect ( as was first understood by Natsuume and Polchinski).
We expect that these additional features will play a central role in full  identification of the bulk dual for the present theory.

\acknowledgments
We acknowledge helpful conversations with Robert de Melo Koch, Gautam Mandal, Pranjal Nayak, Cheng Peng, Kaushik Ray, Al Shapere, Edward Witten and Junggi Yoon on the topics of this paper.
We are especially grateful to Juan Maldacena for several correspondences and discussions.
This work is supported by the Department of Energy under contract DE-SC0010010.
The work of SRD and AG are partially supported by the National Science Foundation grant NSF-PHY-1521045.
The work of KS is also supported by the Galkin Fellowship Award at Brown University.
SRD would like to thank Tata Institute of Fundamental Research and University of Amsterdam for hospitality during the completion of this work.

\appendix
\section{Unit Normalized EAdS/dS Wave Functions}
\label{app:unit normalized eads/ds wave functions}
The unit-normalized Euclidean AdS$_2$ wave function is given by 
	\begin{equation}
		\overline{\phi}_{\rm EAdS_2}(\tau, z) \, = \, \alpha_{\nu} \, z^{\frac{1}{2}} \, e^{-i\omega\tau} \, K_{\nu}(|\omega|z) \, ,
	\label{phi_EAdS}
	\end{equation}
where the normalization factor can be chosen as
	\begin{equation}
		\alpha_{\nu} \, = \, i \, \sqrt{\frac{\nu \sin(\pi \nu)}{\pi^3}} \, .
	\end{equation}
Then from the Bessel $K_{\nu}$ orthogonality condition 
	\begin{equation}
		\int_0^{\infty} \frac{dx}{x} \, K_{i\nu}(x) \, K_{i\nu'}(x) \, = \, \frac{\pi^2}{2} \, \frac{\delta(\nu-\nu')}{\nu\sinh(\pi\nu)} \, ,
	\label{orthogonality of K}
	\end{equation}
where $x, y>0$, the wave function is unit normalized:
	\begin{equation}
		\int_{-\infty}^{\infty} d\tau \int_0^{\infty} \frac{dz}{z^2} \ \overline{\phi}_{\omega, \nu}^{\, *}(\tau, z) \, \overline{\phi}_{\omega', \nu'}(\tau, z) \, = \, \delta(\omega-\omega') \delta(\nu-\nu') \, . 
	\end{equation}
The completeness of modified Bessel function of the second kind:
	\begin{equation}
		\int_0^{\infty} d\nu \, \nu \sinh(\pi \nu) \, K_{i\nu}(x) \, K_{i\nu}(y) \, = \, \frac{\pi^2}{2} \, x \, \delta(x-y) \, ,
	\label{completeness of K}
	\end{equation}
is also used in section \ref{sec:green's function}.

The Lorentzian dS$_2$ wave function is given by 
	\begin{equation}
		\overline{\psi}_{\rm dS_2}(\eta, t) \, = \, \beta_{\nu} \, \eta^{\frac{1}{2}} \, e^{-i\omega t} \, Z_{\nu}(|\omega|\eta) \, .
	\label{psi_dS}
	\end{equation}
Here, let us only consider the continuous modes ($\nu=ir$).
Now choosing the normalization factor as
	\begin{equation}
		\beta_{\nu} \, = \, \sqrt{\frac{\nu}{4\pi \sin(\pi \nu)}} \, ,
	\end{equation}
then the wave function is unit normalized for the continuous modes as
	\begin{equation}
		\int_{-\infty}^{\infty} dt \int_0^{\infty} \frac{d\eta}{\eta^2} \ \overline{\psi}_{\omega, \nu}^{\, *}(\eta, t) \, \overline{\psi}_{\omega', \nu'}(\eta, t)
		\, = \, \delta(\omega-\omega') \delta(\nu-\nu') \, , \qquad {\rm for\ } (\nu=i r)
	\end{equation}
We note that 
	\begin{equation}
		\alpha_{\nu} \, = \, 2i \, \frac{\sin\pi\nu}{\pi} \, \beta_{\nu} \, .
	\label{alpha-beta}
	\end{equation}


\end{document}